\documentclass{llncs}
\usepackage{graphicx}
\usepackage{subfigure}
\usepackage{float}
\usepackage{amsmath}
\usepackage{geometry}

\begin{document}
\title{On Development of Efficient Data Acquisition Systems and Parameter Extraction Technique for DFB Lasers}

%
%
\author{Dao Thanh Hai\inst{1} \and
Le Anh Ngoc \inst{2} \and Ngoc-Cham Vu \inst{2} \and Nguyen Quoc Cuong \inst{3}}
%
%
\institute{Posts and Telecommunications Institute of Technology, Hanoi, Vietnam \and
Electric Power University, Hanoi, Vietnam  \and  Institute of Science and Technology - Ministry of Public Security\\
\email {haidt102@gmail.com}, \email {anhngoc@epu.edu.vn}, \email {chamvn@epu.edu.vn}, \email{cuongnqth@gmail.com}}
\maketitle              
\begin{abstract}
	
Distributed Feedback Laser plays a key role as a light source component in optical fiber communication systems ranging from metro, long-haul to submarine one thanks to its competitive features of superior narrow spectral width and wavelength cohesion. Characterizing such lasers via obtaining their electrical and spectral data and extracting their internal parameters therefore remains a critical task in designing and troubleshooting optical fiber systems. This paper presents first an agile framework for a rapid collection of laser data via automatic measurement and second an efficient approach for extracting laser internal parameters.

\end{abstract}
\section{Introduction}
Internet has become clearly the largest engineered system made by humankind with millions of end devices, telecommunication links, switches and routers connecting to each others and billions of users have been on Internet via different means \cite{hai_oft}, \cite{hai_access}, \cite{hai_comcom}, \cite{hai_comcom2}, \cite{hai_springer}. One one hand, such exponential growth of Internet users is driven by the rise of non-conventional applications including Internet of (every-) things (IoTs), machine-to-machine connections (M2M). On the other hand, more devices connecting to the Internet implies that there will be a severe impact on Internet traffic. Indeed, globally, Internet traffic will grow 3.7-fold from 2017 to 2022, a compound annual growth rate of $30\%$ and in parallel, the significance advances in telecommunication infrastructure, including access, metro and backbone networks will be expected. It has to be noted that several billion kilometers of optical fiber have been installed around the globe today, constituting the major segment of Internet infrastructure \cite{hai_oft21}, \cite{hai_iet}, \cite{hai_wiley}, \cite{hai_optik19}. In essence, it is important to recognize that for every bit of information we send or receive over Internet today, the major part of its journey in the form of photon traversing via global optical networks infrastructure \cite{hai_sigtel1}, \cite{hai_sigtel2}, \cite{hai_csndsp}, \cite{hai_ps1}, \cite{hai_ps2}, \cite{hai_icist1}

Laser is a special device which is capable of emitting light through the optical amplification process relied on a physical phenomenon, called, stimulated emission of electromagnetic radiation. For a laser to be operational, two conditions have to be available, that is, the cavity and a gain medium in that cavity and the type of laser is indeed determined by the type of gain medium. Some of the lasers have been widely used in practice for various purposes including Fabry-Perot (FP) lasers, distributed feedback (DFB) lasers, external cavity lasers (ECLs) \cite{dfb2}. In this paper, we focus on DFB laser as this is the most widely used one for high-speed optical networks and its development is indeed closely tied to the advances in both speed and reliability of the next-generation optical fiber systems. By controlling the period of the Bragg grating, the center peak wavelength tunability of DFB lasers could be realized and such important property is critical in telecommunication systems. In practice, such tunability is controlled by adjusting the device temperature and DFB laser can therefore work smoothly over a whole C-band in optical systems \cite{dfb3}.
 
This paper reports our work on developing an efficient data acquisition system for automatic measuring of key physical properties of DFB lasers. Such system is designed and programmed in Labview as described in Sect. \ref{sect: labview}. In processing the collected data to characterize laser as detailed in Sect. \ref{sect: dataprocessing}, we develop circuit models to extract physical parameters. Finally, the paper is concluded with some remarks to summarize and open up possibly follow-up works. 

\section{Labview Programs for Automatic Data Acquisition System}
\label{sect: labview}
\subsection{$P, V, I_{PD}$ Acquisition} 
\begin{figure}[!h]
	\centering
	\includegraphics[scale=0.8]{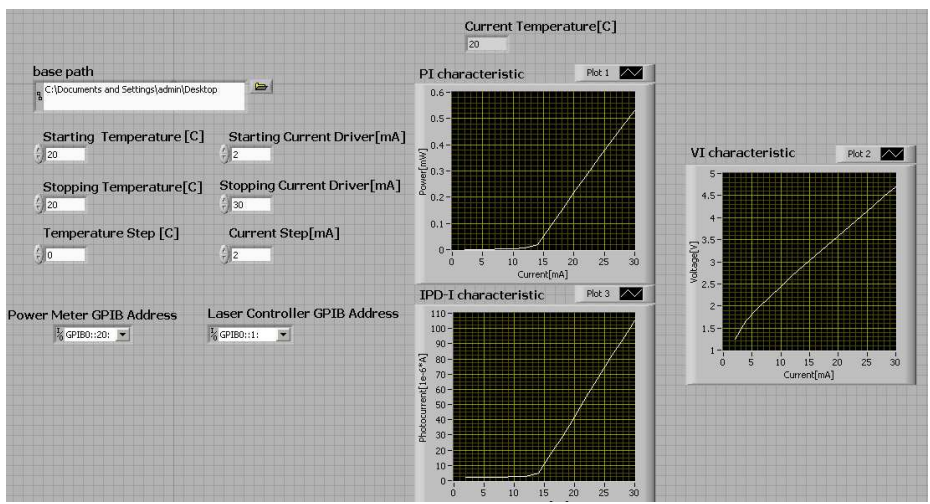}
	\caption{VI to collect $P,V,I_{PD}$ over a range of $T^{o}$ and $I$}
	\label{fig:DAQ_1}
\end{figure}

This Labview program enables users to observe the three important dependencies, namely, $P-I$, $V-I$ and $I_{PD}-I$ at different range of temperature and current. The user can either opt for data collection and visualization at a specific point\textemdash designated input of laser temperature and current\textemdash or in a batch mode in an automatic manner over several data points. Besides, our Labview program allows real-time visualization of such measurement in three diagrams as shown in Fig. \ref{fig:DAQ_1}. Moreover, the data file could be exported to predefined directory according to structure in Tab. \ref{tab:DAQ_1}.

\begin{table}[!h]
	\centering
	\caption{Data Description for first VI}
	\begin{tabular}{cc}
		\hline
		& Output Files Description\\ \hline
		File Name & PI\_Temperature\_Start Current\_Stop Current\_Current Step.txt \\[0.5ex]
		& VI\_Temperature\_Start Current\_Stop Current\_Current Step.txt \\
		& IPDI\_Temperature\_Start Current\_Stop Current\_Current Step.txt \\  \hline
		Data Organization & $PI$:$1^{st}$column is Power[W]; $2^{nd}$ column is Current[mA]\\
		&$VI$: $1^{st}$ column is Voltage[V]; $2^{nd}$ column is Current[mA]\\
		&$I_{PD}I$: $1^{st}$ column is Photocurrent[$\mu$A]; $2^{nd}$ column is Current[mA]\\ 
		\hline
	\end{tabular}
	
	\label{tab:DAQ_1}
\end{table}

\subsection{Spectrum Acquisition} 
\begin{figure}[!h]
	\centering
	\includegraphics[scale=0.8]{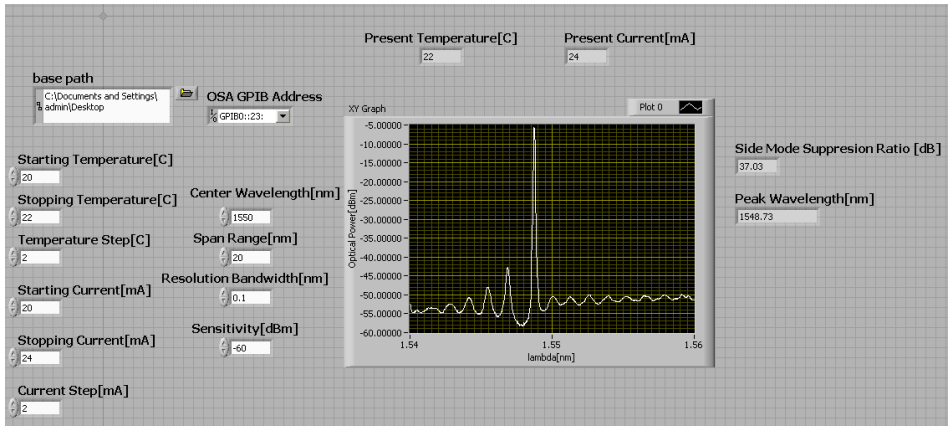}
	\caption{VI to collect spectrum over a range of  $T^{o}$ and $I$ }
	\label{fig:DAQ_2}
\end{figure}

This program, shown in Fig. \ref{fig:DAQ_2}, is developed to offer automatic feature on spectrum data collection. In addition to the (almost) real-time visualization of measured data together with the display of key measured information including peak wavelength and SMSR, our program allows the exportation of all raw measurement data. Table \ref{tab:DAQ_2} provides descriptions of how data is collected and stored in this Labview program.

\begin{table}[!h]
	\centering
	\caption{Data Description for second VI}
	\begin{tabular}{cc}
		\hline
		& Output Files Description\\ \hline
		File Name & OSA\_Temperature\_Current.txt \\
		&peaklambda\_start temp\_stop temp\_start current\_stop current.txt \\
		&SMSR\_start temp\_stop temp\_start current\_stop current.txt \\ \hline
		Data Organization & $OSA$:$1^{st}$column is $\lambda[nm]$; $2^{nd}$ column is Power[dBm]\\
		&$peaklambda$: Columns and rows correspond to $T^{o}, I$; the unit is [nm]. \\
		&$SMSR$: Columns and rows correspond to $T^{o}, I$; the unit is [dB] \\  
		\hline
	\end{tabular}
	
	\label{tab:DAQ_2}
\end{table}

\section{Data Processing}
\label{sect: dataprocessing}
Electrical and Spectral Data of our laser (DFB) are processed in order first to characterize laser and seconnd to extract internal parameters. The work here involves finding threshold current, and through that evaluating the characteristic temperature $T_0$, investigating the dependence of output characteristics ($P$ and$\lambda _p$) on input parameters ($T^0$ and $I$), as well as estimating the coupling coefficient.

\begin{figure}[!htbp]
	\centering
	\begin{minipage}[b]{0.5\textwidth}
		\includegraphics[width=\linewidth]{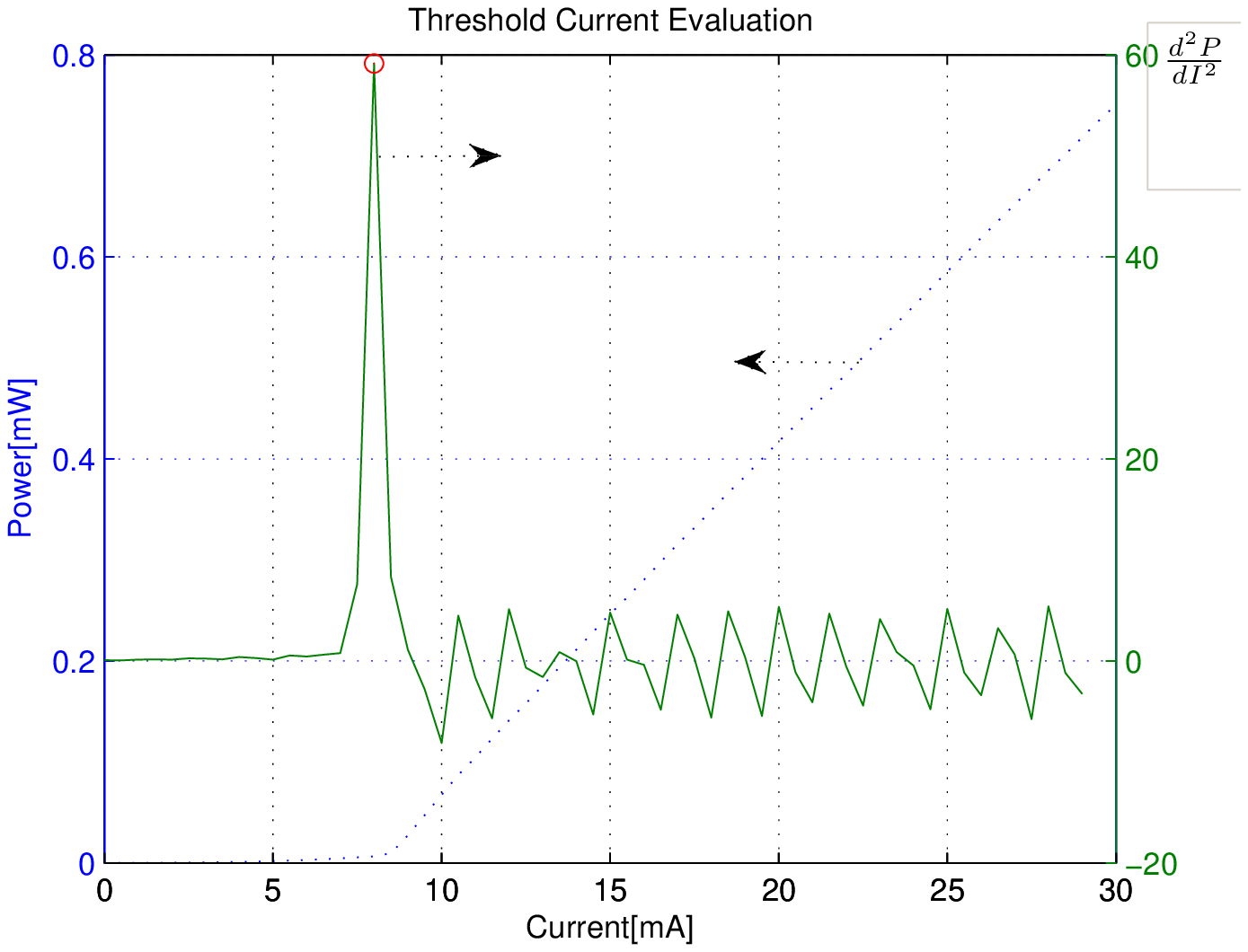}
		\caption{Illustration of Threshold Calculation}
		\label{fig:threshold}
	\end{minipage}%
	\hfill
	\begin{minipage}[b]{0.5\textwidth}
		\includegraphics[width=\linewidth]{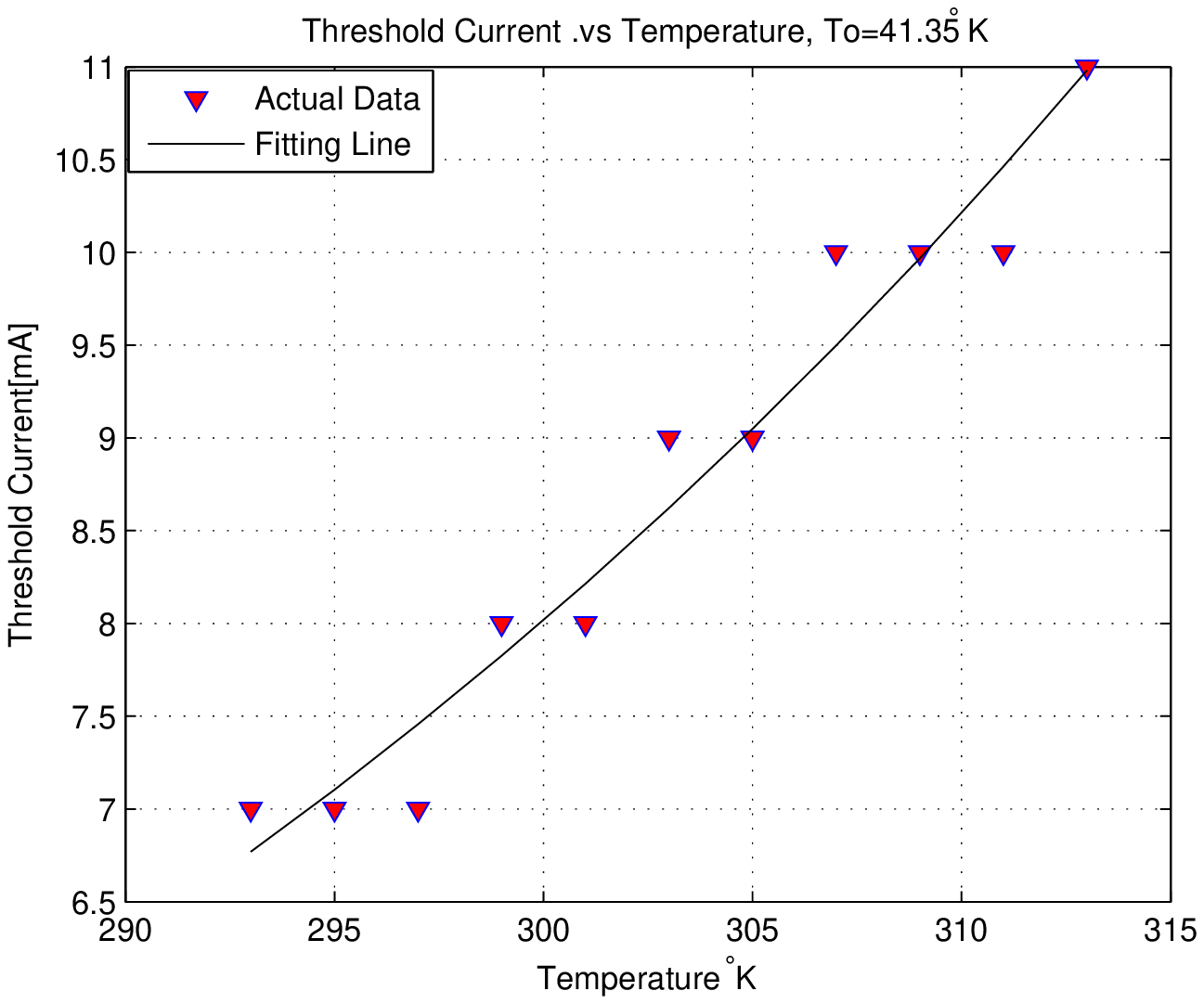}
		\caption{$T_0$ evaluation}
		\label{fig:To}
	\end{minipage}
\end{figure}

There are number of ways to extract threshold current from P-I characteristic. In \cite{dfb1}, \cite{dfb3}, the methods as well as their pros and cons are discussed in-depth. The approach based on looking for maximum of second derivative of P over I is reported here thanks to its superior tolerance to non-linearities before and after threshold knee. Illustration for this method at $T=25^o C$ is shown in Fig. \ref{fig:threshold}. The characteristic temperature is subsequently derived based on the model suggested in \cite{dfb3}, that is $I_{th}=I_0e^{T/T_0}$. The entire results are shown in Fig. \ref{fig:threshold} and Fig. \ref{fig:To}

\begin{figure}[!h]
	\centering
	\includegraphics[scale=0.5]{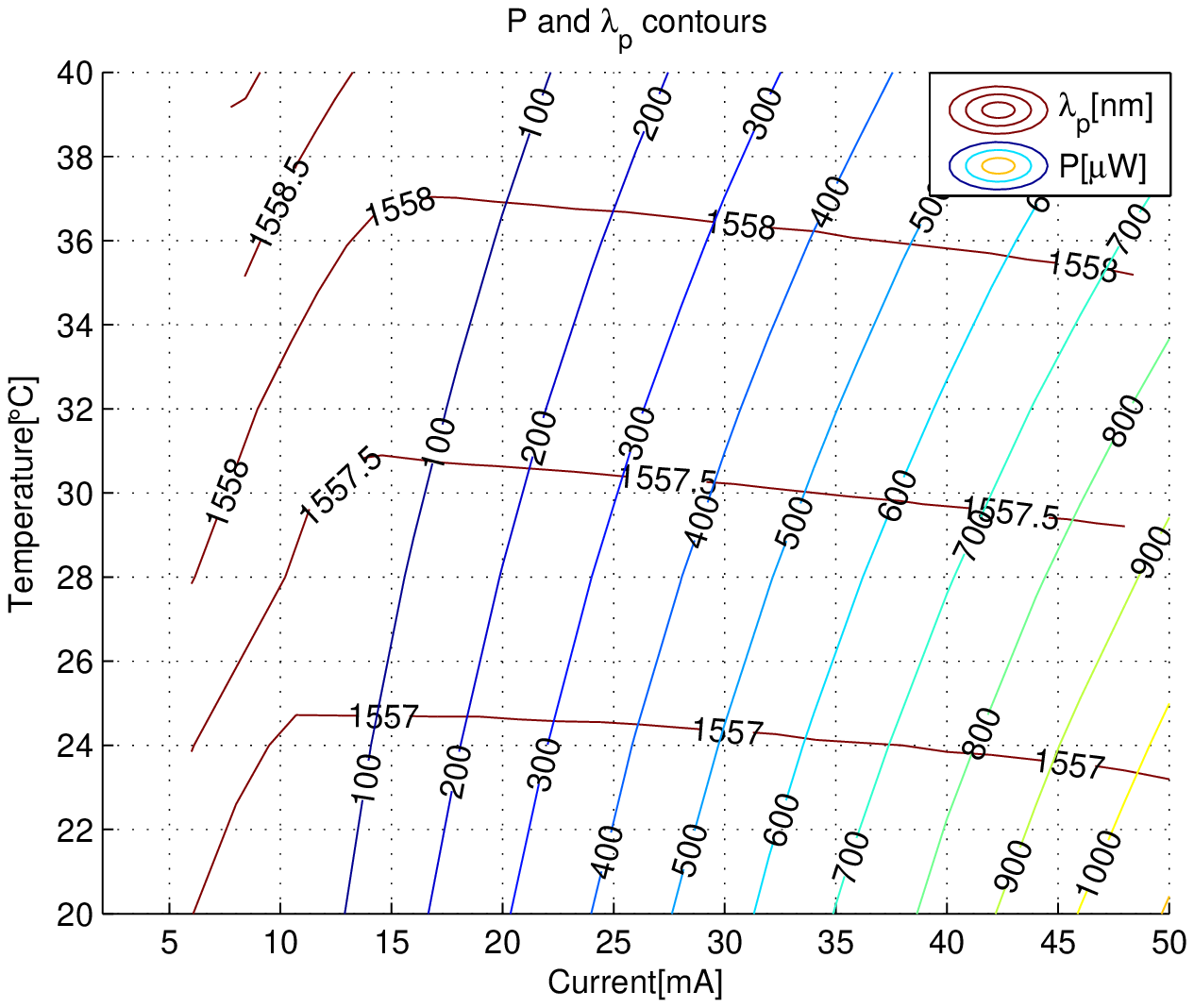}
	\caption{$P$ and $\lambda$ contours}
	\label{fig:Userguide}
\end{figure}

Peak Wavelength and Emitted power are both influenced and/or shifted by whatever change in drive current and temperature. Figure. \ref{fig:Userguide} serves as  graphical map to observe this variation. In certain sense, that map may assist users in adjusting input parameters to reach desired outputs 

\textbf{Coupling Coefficient} is important information for optimization of DFB laser as well as for system design purposes and therefore finding ways for effective estimation is still currently topic of interest. In general, $\kappa$ is different at forward ($\kappa_{RS}$) and backward ($\kappa_{SR}$) direction. Its mathematical expression is: $\kappa= \kappa_{i}+j\kappa_{g}e^{\pm \theta}$ (see \cite{dfb3}).

\begin{figure}[!tbp]
	\centering
	\begin{minipage}[b]{0.5\textwidth}
		\includegraphics[width=\linewidth]{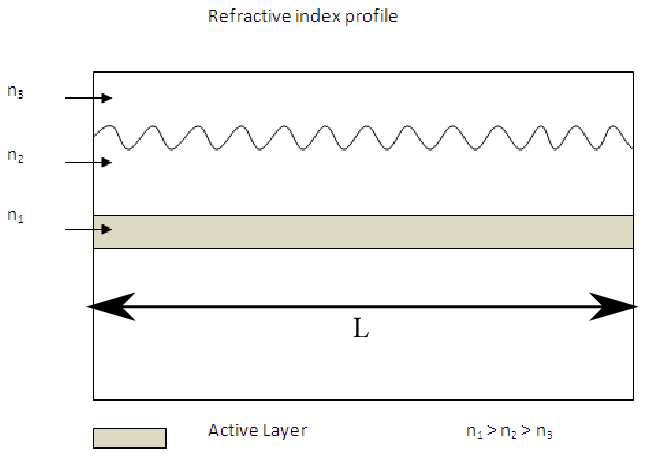}
		\caption{index-coupled DFB laser}
		\label{fig:couplingcoeff1}
	\end{minipage}%
	\hfill
   \begin{minipage}[b]{0.5\textwidth}
   	\includegraphics[width=\linewidth]{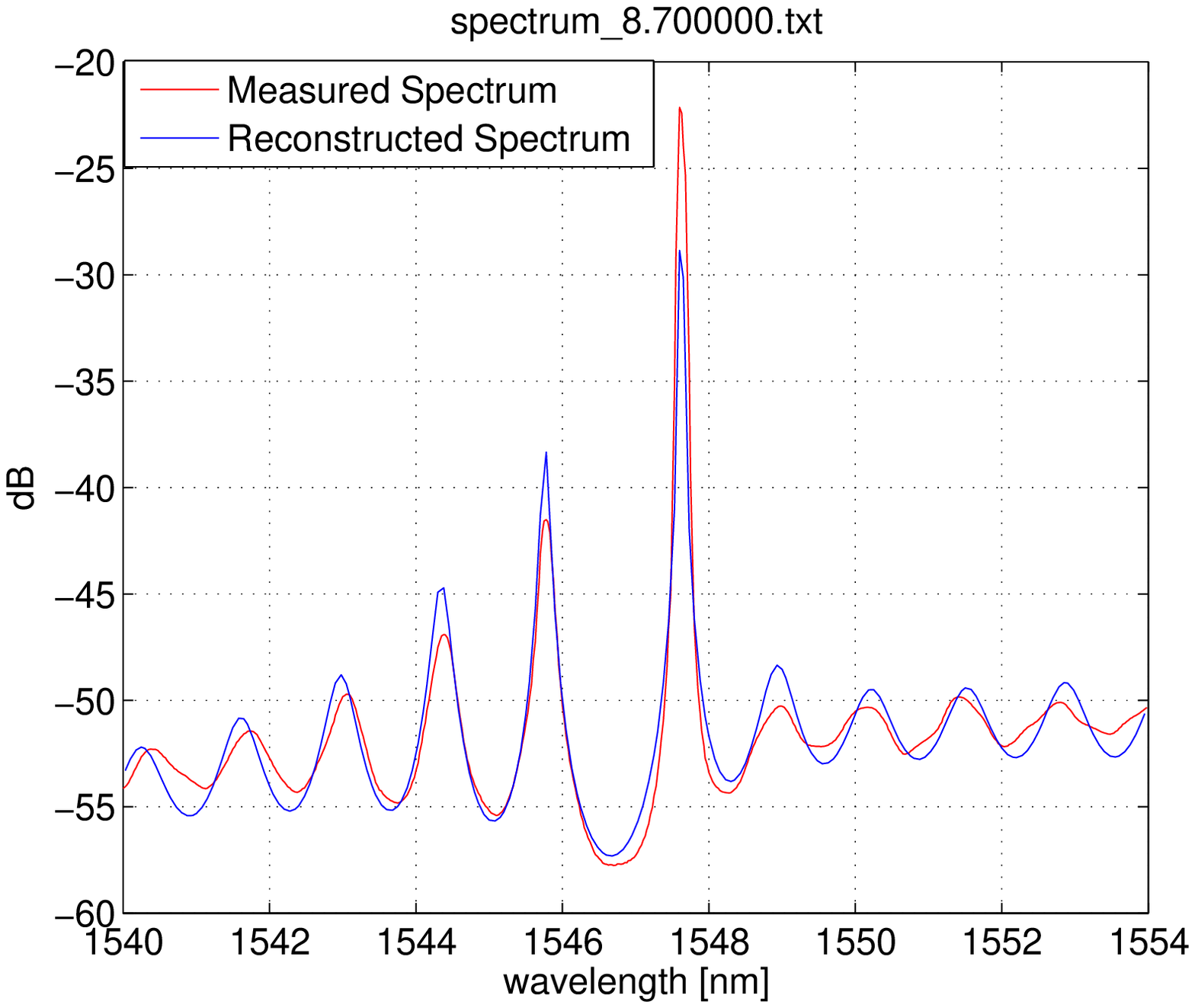}
   	\caption{Coupling Coefficient Extraction}
   	\label{fig:couplingcoeff2}
   \end{minipage}
\end{figure}

Our approach to evaluate $\kappa$ is based on the method suggested in \cite{dfb3}, that is to use numerical fitting of theoretical sub-threshold spectrum into measured one by least-square algorithm. For simplicity, we assume our laser belong to purely index-coupled one (see Fig. \ref{fig:couplingcoeff1}), which is commonly the case. Under this assumption, only refractive index varies along the longitudinal direction and hence we have $\kappa_{RS}=\kappa_{SR}=\kappa_{i}=\kappa $. The outcome of the work gives us $\kappa_{i}L=0.66$

\section{Conclusions}
To sum up, two main contribution have been reported in this paper, that is to build tools supporting automatic data collection and to employ parameters extraction techniques for laser characterization. Our tools proposed in this paper allows an efficient and agile procedure with low-complexity to rapidly measure both electrical and spectral data of DFB lasers, which are particularly helpful in the realm of laser measurements and laser modeling. \\

\footnotesize{\textbf{Acknowledgement}: This work is funded by the Vietnam National Foundation for Science and Technology Development (NAFOSTED) under the grant number 102.02-2018.09}

 \bibliographystyle{wileyj}
 \bibliography{ref}

\end{document}